\documentclass[a4paper,conference]{IEEEtran}

\ifCLASSINFOpdf
   \usepackage[pdftex]{graphicx}
   \graphicspath{{./images/}}
   \DeclareGraphicsExtensions{.pdf,.jpeg,.png}
\else
   \usepackage[dvips]{graphicx}
   \graphicspath{{./images/}}
   \DeclareGraphicsExtensions{.eps}
\fi

\ifCLASSOPTIONcompsoc
  \usepackage[caption=false,font=normalsize,labelfont=sf,textfont=sf]{subfig}
\else
  \usepackage[caption=false,font=footnotesize]{subfig}
\fi

\newcommand{\EMOGAS}{$SpO_{2}/FiO_{2}$}

\usepackage{multirow}
\usepackage{color}
\usepackage{url}
\begin{document}
\title{Deep learning in the ultrasound evaluation of neonatal respiratory status}
\author{\IEEEauthorblockN{Michela Gravina\IEEEauthorrefmark{1},
Diego Gragnaniello\IEEEauthorrefmark{1},
Luisa Verdoliva\IEEEauthorrefmark{2},
Giovanni Poggi\IEEEauthorrefmark{1},
Iuri Corsini\IEEEauthorrefmark{4},
Carlo Dani\IEEEauthorrefmark{4},\\
Fabio Meneghin\IEEEauthorrefmark{5},
Gianluca Lista\IEEEauthorrefmark{5},
Salvatore Aversa\IEEEauthorrefmark{6},
Francesco Raimondi\IEEEauthorrefmark{3},
Fiorella Migliaro\IEEEauthorrefmark{3} and
Carlo Sansone\IEEEauthorrefmark{1}
}
\IEEEauthorblockA{\IEEEauthorrefmark{1}Dept. of Electrical Eng. and Information Technology, University of Naples Federico II, Naples, Italy}
\IEEEauthorblockA{\IEEEauthorrefmark{2}Dept. of Industrial Engineering, University of Naples Federico II, Naples, Italy}
\IEEEauthorblockA{\IEEEauthorrefmark{3}Division of Neonatology, Section of Pediatrics, Dept. of Translational Medical Sciences,\\University of Naples Federico II, Naples, Italy}
\IEEEauthorblockA{\IEEEauthorrefmark{4}Hospital Careggi of Florence, Italy}
\IEEEauthorblockA{\IEEEauthorrefmark{5}Hospital “Vittore Buzzi”, Milan, Italy}
\IEEEauthorblockA{\IEEEauthorrefmark{6}Civil Hospital of Brescia, Italy}
}

\maketitle

\begin{abstract}
Lung ultrasound imaging is reaching growing interest from the scientific community.
On one side, thanks to its harmlessness and high descriptive power, this kind of diagnostic imaging has been largely adopted in sensitive applications, like the diagnosis and follow-up of preterm newborns in neonatal intensive care units.
On the other side, state-of-the-art image analysis and pattern recognition approaches have recently proven their ability to fully exploit the rich information contained in these data, 
making them attractive for the research community.
In this work, we present a thorough analysis of recent deep learning networks and training strategies carried out on a vast and challenging multicenter dataset comprising 87 patients with different diseases and gestational ages.
These approaches are employed to assess the lung respiratory status from ultrasound images and are evaluated against a reference marker.
The conducted analysis sheds some light on this problem by showing the critical points that can mislead the training procedure and proposes some adaptations to the specific data and task.
The achieved results sensibly outperform those obtained by a previous work, which is based on textural features, and narrow the gap with the visual score predicted by the human experts.
\end{abstract}

\IEEEpeerreviewmaketitle

\section{Introduction}
The clinical study of lung disease of the adult has its gold standard in the Chest X-Ray (CXR).
However, when resorting to preterm newborns,
CXR results in poor prognostic capabilities.
Several studies have pointed out the limited performance into predicting either the need for respiratory support \cite{dimitriou1995appearance} or the rise of chronic lung diseases \cite{greenough2004prediction}.
Moreover, they observed a low correlation between radiological scores and the respiratory function,
with no inter-observer agreement \cite{snepvangers2004chest}.
Above all, such a poor prediction capability comes at the cost of a very high radiological risk that the patient can accumulate for the rest of his/her life.

On the contrary, lung ultrasound (US) has gained more attention in recent years
because it is harmless and exhibits a high descriptive power.
With its ability to analyze both anatomical structures, like the pleural line,
and artifacts, related to the lung composition and status,
the US images is prone to many investigations through either visual inspection or computer-aided analysis.
Early studies presented detailed procedures to visual assess the respiratory status of newborn patients \cite{raimondi2013point}, as well as the need for respiratory support \cite{brat2015lung}.
Recently, the activity in this research area is boosted by the availability of powerful image processing tools, in particular those based on deep learning, able at solving complex tasks about image analysis.
Recent studies have focused their attention on fetal lung US images
and proposed new approaches to assess the respiratory status \cite{perez2019clinical,du2020application}
or to predict the newborn diseases \cite{abdelhamid2020quantitative}.
In this work, we collected a large and challenging dataset and tested different deep convolutional neural networks.
The proposed training strategies to cope with this particular task and specific kind of data,
show a performance improvement that narrows the gap between fully automatic tools and human experts.

The rest of this paper is organized as follows.
In Sec. \ref{sec:related_works}, related works are discussed, while
the collected dataset is described in Sec. \ref{sec:dataset}.
In Sec. \ref{sec:exp_setting}, the relation between data augmentation for different training strategies and the considered problem are discussed and some advanced approaches are presented.
%in order to tailor the considered deep learning techniques to this specific problem.
In Sec. \ref{sec:results}, both quantitative and qualitative results are presented.
Finally, conclusions are drawn in Sec. \ref{sec:conclusions}.

\section{Related Works}
\label{sec:related_works}

Different works published in the scientific literature present semi- or fully-automatic approaches
to assess the health status of adult patients through the analysis of the lung US images.
Among the first ones, some works have explored the use of computer-aided analysis based on the first-order statistics \cite{corradi2015quantitative, corradi2016computer}.
However, the recent advent of deep learning techniques has given rise to many novel approaches.
In particular, the attention on lung US automatic analysis has experienced a rapid growth due to the COVID-19 pandemic \cite{sultan2020review}.
As an example, in \cite{born2020pocovid}
a modified version of VGG has been proposed,
while an approach based on Radon transform able at detecting LUS B-lines has been presented in \cite{karakucs2020line} and in \cite{anantrasirichai2017line} for general lung health assessment.
In \cite{roy2020deep},
the authors presented an approach to jointly detect and localize COVID-19 in lung US.
After a Spatial Transformer Networks \cite{jaderberg2015spatial}, each frame is classified and a video-level score aggregation has been performed through uninorm operators \cite{yager1996uninorm}. 
Segmentation is addressed via a dedicated network trained in a fully-supervised (pixel-wise) fashion.

Research on preterm newborns has fewer contributions, probably due to the scarcity of data.
Among the most recently proposed approaches,
several studies have focused their attention on fetal lung US image analysis.
These methods, mostly based on very well-known textural descriptors,
aimed to assess the respiratory status \cite{perez2019clinical,du2020application} or predict the newborn diseases \cite{abdelhamid2020quantitative}.
Texture analysis has been also employed to determine the respiratory morbidity in newborns \cite{palacio2017prediction}, while
recent fully automatic methods focused on predicting the neonatal maturation degree \cite{chen2020preliminary} as well as the respiratory morbidity \cite{burgos2019evaluation}.

In our preliminary work \cite{raimondi2018visual}, we presented a first attempt to define a score,
directly obtained from US images,
and test its correlation with two oxygenation indexes, namely the oxygenation ratio ($PaO_{2}/FiO_{2}$) and the alveolar arterial oxygen gradient ($A-a$), both considered as reference standards.
A significant correlation was found both with the proposed visual scores, namely the LUS score, assigned by two different experts, and the automatic score obtained by the texture analysis of the acquired digital US videos.
So far, the advances in assessing the respiratory status of preterm newborns are scarce.
In this work, we aim to extend the experimental analysis by largely increasing the dataset size,
which now includes videos acquired by three centers and patients with a significant gestational age difference and eventually affected by two diseases.
Differently from our previous work, we selected a non-invasive reference standard, i.e. the pulse oximetric saturation ratio \EMOGAS~(SF) value, so as to make the whole process non-invasive.
Finally, we analyze and assess the performance of recent state-of-the-art Convolutional Neural Networks (CNNs) and training strategies for the task of predicting the lung functionality status in preterm babies.

\section{Dataset}
\label{sec:dataset}
\begin{figure*}[!t]
	\centering
	\subfloat[][Healthy (Naples)]
	{\includegraphics[width=.3\linewidth]{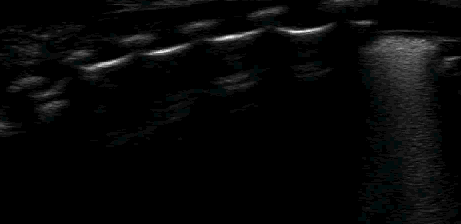} \label{fig:bestNapoli}}
	\subfloat[][TTN (Naples)]
	{\includegraphics[width=.3\linewidth]{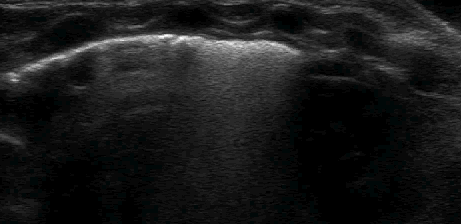} \label{fig:ttnNapoli}}
	\subfloat[][RDS (Naples)]
	{\includegraphics[width=.3\linewidth]{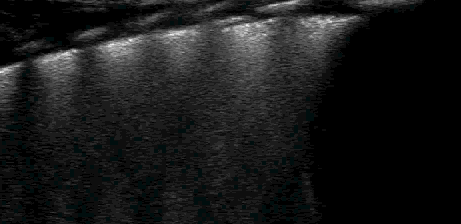} \label{fig:rdsNapoli}}
	
	\subfloat[][Healthy (Florence)]
	{\includegraphics[width=.3\linewidth]{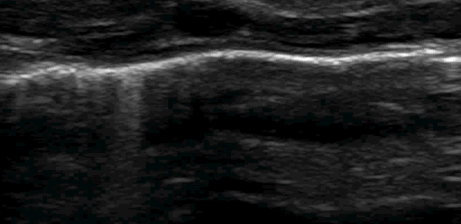} \label{fig:bestFirenze}}
	\subfloat[][TTN (Naples)]
	{\includegraphics[width=.3\linewidth]{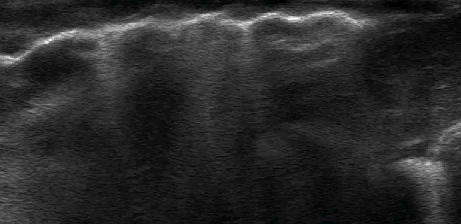} \label{fig:ttnNapoli2}}    
	\subfloat[][RDS (Florence)]
	{\includegraphics[width=.3\linewidth]{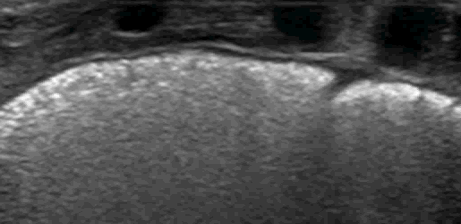} \label{fig:rdsFirenze}}
	
	\subfloat[][Healthy (Milan)]
	{\includegraphics[width=.3\linewidth]{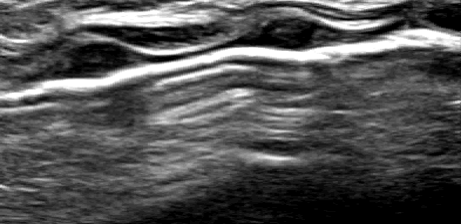} \label{fig:bestMilano}}
	\subfloat[][TTN (Naples)]
	{\includegraphics[width=.3\linewidth]{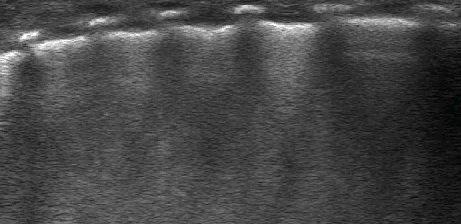} \label{fig:ttnNapoli3}}    
	\subfloat[][RDS (Milan)]
	{\includegraphics[width=.3\linewidth]{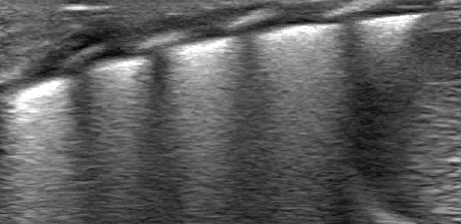} \label{fig:rdsMilano}}
	
	\caption{From left to right: healthy, TTN and RDS patients. When available, each row depicts patients from a different center.}
	\label{fig:SchemaImag}
\end{figure*}

The dataset consists of US videos of 86 patients from 3 centers with different respiratory diseases,
namely Respiratory Distress Syndrome (RDS) and Transient Tachypnea of the Newborn (TTN).
Their gestational age ranges from 25 to 40 weeks.
Videos are acquired with several US instruments, yielding to different resolutions and frame rates.
Figure \ref{fig:SchemaImag} shows examples of frames acquired by all three centers and depicting different respiratory diseases.
All the videos acquired at the same time from different views of both patient lungs are grouped in a session,
which is associated with the pulse oximetric saturation ratio (SF) value, the marker we selected as golden standard.
It should be noted that this marker, which can be easily measured by means of a pulse oximeter,
is non-invasive, unlike the oxygenation ratio and the alveolar arterial oxygen gradient that are obtained through blood collection.
From each video 6 up to 10 evenly spaced frames are selected.
During the performance evaluation, a maximum of 6 frames are considered for each video,
so as to neglect the effect of different video lengths or frame rates.
Healthy patients underwent US only once, meanwhile the others had 2 or 3 US sessions,
the last of which acquired after the patient has been healed.
Details of the dataset are reported in Table \ref{tab:dataset}.

\begin{table}[!t]
\centering
\caption{The collected dataset}
\label{tab:dataset}
\begin{tabular}{|l|cc||c|c|c|}
	\hline
	Disease & \multicolumn{2}{c||}{Number of} & \multicolumn{3}{c|}{Patients per center} \\
	        & patient &        videos         & Naples & Milan &        Florence         \\ \hline
	None    &   43    &         ~328          &   23   &  10   &           10            \\ \hline
	RDS     &   30    &         ~727          &   11   &   8   &           11            \\ \hline
	TTN     &   13    &         ~211          &   13   &   -   &            -            \\ \hline\hline
	Tot.    &   86    &         1266          &   47   &  18   &           21            \\ \hline
\end{tabular}
\end{table} 

\section{Experimental setting}
\label{sec:exp_setting}

\subsection{Pre-processing and Data augmentation}

Firstly, the extracted frames are normalized in the 0-1 range and resized to have the same horizontal resolution. For the majority of the dataset videos, this is equal to 461, thus we decided to keep this value. After that, only the first $R$ rows, where $R$ is the standard resolution of the considered network, are retained.
For all of the videos and values of $R$, this always includes the pleural line and part of the lung below it.
Such a rectangular-shaped ($R\times461$) input allows the network to process the whole US from left to right as done by the human experts without discarding any important detail.
At the same time, we neglect the bottom region of the US images, which carries no useful information.

Given the very particular images we are dealing with, we carefully selected the plausible data augmentations.
Random horizontal flip is always adopted in our experiments.
On the contrary, a vertical flip would yield to a non-sense image where the pleural line is upside-down, thus this operation is not considered.
Differently from other medical imaging procedures, US images require a manual and time-varying process especially for preterm newborns, where a human operator carefully handles a probe 
to perform the chest scan from different views.
Hence the final result is affected by the hand movement, in terms of both speed and stability, and by the frame acquisition rate.
In this respect the simplest data augmentation is to select more frames from each video.
During training, at most 10 evenly spaced frames are selected from each video,
while at test time only 6 frames are taken.
This allows us to select quite distant/diverse frames and, at the same time, to select the same number of samples for each video no matter its length.
Beyond this, we also randomly rotated the image in the range $\left[-10^\circ, 10^\circ\right]$ in order to simulate different incidence angles of the probe. A stronger rotation would give an unnatural image, thus it is avoided.

Finally, we deal with different probes and acquisition instruments that could be calibrated with dissimilar sensitivity with respect to each other.
Indeed, after the whole acquisition process the very same ultrasound wave can be converted in two different gray scales, e.g. brighter for more sensible instruments and darker for the less sensible ones.
To simulate multiple acquisitions of the same image with multiple devices, and thus make the training procedure robust against different calibrations and instruments,
we randomly modified the brightness and the contrast of the images in a relative range of $25\%$.

\subsection{Convolutional Neural Networks}
We considered five different networks.
The first one, AlexNet \cite{Alexnet2012}, is a very well-known CNN used as a baseline.
The second one is a ResNet \cite{ResNet34}, another very famous CNN architecture.
Thanks to its peculiar skip connections, this network can easily be trained
avoiding the vanishing gradient problem and, at the same time, making the architecture more efficient.
After a preliminary analysis, we selected the model with 34 layers that represents a good compromise between complexity and descriptive power.
On the same path there is EfficientNet \cite{tan2019efficientnet}, a recently proposed class of scalable CNN architectures.
These networks differ from each other with respect to a scaling factor that modifies both the depth, width, and resolution of the architectures.
Among these, we selected those with a number of training parameters more similar to that of ResNet, i.e. EfficientNet-B0, -B1 and -B2.

\subsection{Baseline training strategies}
Deep convolutional neural networks (CNNs) are robust and powerful tools that can be trained with different targets and tasks.
In this work, we pursued two different training strategies to achieve our goal, i.e. predict a score that correlates with the SR marker.

The first and more direct approach is to train the CNN with a regression loss, e.g. mean squared error, to directly predict the SR value,
eventually normalized in the unity range for better training stability.
To this aim, the SF values should be available during training.
However, this is not a real limitation considering that the SF marker is non-invasive and can be easily measured at any time.
For this training strategy, distinguishing between two very high values (e.g. SF above 450),
which represent either an healthy or a completely healed patient,
is useless.
To avoid focusing the training phase on these unessential differences, we eventually clipped the SF value to 450.
Performance are evaluated on both the Spearman's rank correlation coefficient and the Mean Absolute Percentage of Error (MAPE), defined as:
\begin{equation}
MAPE(x,y) = \frac{1}{N}\sum_n\frac{\left|x_n-y_n\right|}{y_n},
\end{equation}
where $x_n$ and $y_n$ are the predicted and the target value, and $N$ is the number of samples in the test set.
Values are aggregated at video-level and session-level by averaging the frame-level predictions.

Another approach that can be pursued is to train the network with class labels, i.e. healthy or sick,
that will be predicted with a certain confidence.
When normalized in the unit range, such a confidence score can be regarded as a class probability,
which we can correlate with the SF.
Indeed, the more serious is the pathology, the lower will be the SF value.
Differently from the regression-based training, in this alternative scenario, the actual SF values are never used by the network during training.
Performance at both frame-, video- and session-level are evaluated on both the Spearman's rank correlation coefficient and the Accuracy.

\subsection{Advanced training strategies}
\begin{figure}[!t]
	\centering
	\includegraphics[width=3.5in]{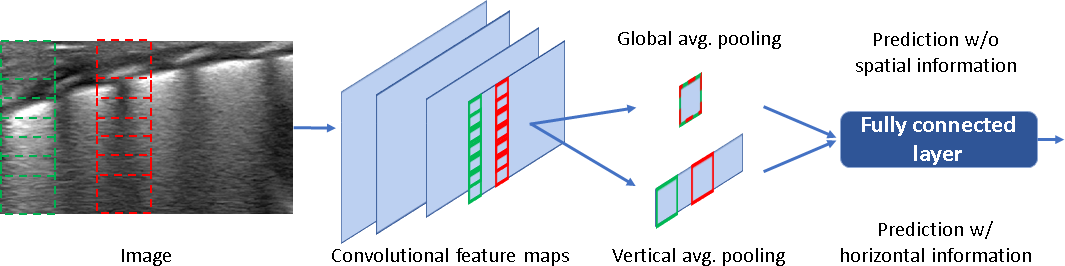}
	\caption{Scheme of the proposed vertical average pooling. This strategy preserves the horizontal position of the extracted features at the cost of an higher number of training parameters.}
	\label{fig:scheme}
\end{figure}

To better adapt the networks to this specific problem we propose two different approaches.
From one side, we tried to preserve the horizontal position of the extracted features.
This is motivated by the US imaging process.
Normally, each vertical stripe of the lung US shows a part of the pleural line and the lung below it, both containing useful information for our problem.
However, when the US wave meets an obstruction (e.g. a rib), the resulting vertical stripe of the image shows no signal (black).
Unfortunately, this can mislead the model training because the healthy lung US is characterized by the prevalence of dark regions.
To avoid this, we replace the global average pooling,
often present in recent CNN architectures between the last convolutional layer and the final classification stage,
with a row-wise average pooling that preserves the horizontal flip position of each feature and
avoids mixing feature maps extracted by different vertical stripes of the image. A scheme representing this strategy is depicted in Fig. \ref{fig:scheme}.

The severity of the disease is another information that we can provide to the model during training.
While the regression-based training strategy implicitly receive such information directly from the target SF value to predict,
as regarding the classification-based training strategy we resort to the use of the curriculum learning \cite{curriculum}.
This training procedure consists of two phases: i) the network is trained with \emph{easy} samples, and
ii) \emph{hard or borderline} samples are added to the training set.
This method has proven superior performance and faster convergence in different medical applications \cite{tang2018attention}.

\section{Results}
\label{sec:results}
We carried out a preliminary analysis considering the ResNet architecture to select a proper depth and image crop. Results are reported in Tab. \ref{tab:preliminary}.
As regarding the network depth, ResNet18 showed interesting results in terms of accuracy.
However, as regarding the correlation with the marker, our main goal, ResNet34 outperformed the others.
For this particular model we also report the performance obtained adopting a random crop augmentation, where a squared patch lying in the selected 224$\times$461 pixels region of interest is randomly selected.
In this case, the performance drop confirms that the whole LUS should be analyzed by the network at the same time.

\begin{table}[!t]
	\renewcommand{\arraystretch}{1.3}
	\caption{Preliminary experiments carried out using the ResNet architecture by varying the depth and the crop strategy. Both the frame-level correlation with the reference marker and the accuracy are reported.}
	\label{tab:preliminary}
	\centering
		\begin{tabular}{|llc||c|c|}
		\hline
		Depth  & Crop        &   Input size   &  Correlation    &  Accuracy       \\\hline\hline
		18     & Whole width & 224$\times$461 &     0.6411      & \textbf{0.8221} \\ \hline
		34     & Whole width & 224$\times$461 & \textbf{0.6507} &     0.8129      \\ \hline
		50     & Whole width & 224$\times$461 &     0.6283      &     0.8080      \\ \hline
		34     & Random      & 224$\times$224 &     0.6429      &     0.8198      \\ \hline
	\end{tabular}
\end{table}

The results obtained for all the networks by varying data augmentation are reported in Tab. \ref{tab:baselines_reg} and Tab. \ref{tab:baselines_class}
for regression and classification baseline training strategies, respectively.
From top to bottom, a stronger augmentation often improves performance, with some exceptions.
For example, the regression problem (Tab. \ref{tab:baselines_reg}) seems more difficult to address,
and the networks are not able to take full advantage of a strong augmentation.
In particular, when no or a light augmentation is performed (Tab. \ref{tab:baselines_reg}, top),
EfficientNet-B1 shows nice performance, achieving the best results in terms of MAPE and video-level correlation.
For what concerns the correlation metric, best performance at the frame- and session-level are respectively obtained with AlexNet (strong augmentation) and ResNet (mid augmentation).
However, the gap with the runner up is tiny in this case.

\begin{table*}[!t]
	\renewcommand{\arraystretch}{1.3}
	\caption{Regression results of the considered baselines with different augmentations. From top to bottom we report the results without any augmentation and by adding the following operations, consecutively: random horizontal flip, random rotation in the range $\left[-10^\circ, +10^\circ\right]$, random brightness and or contrast variation in the range $\left[-25\%, +25\%\right]$.}
	\label{tab:baselines_reg}
	\centering
	\begin{tabular}{|lc|c||ccc|ccc|}
		\hline
		Network         &   Input size   &       Augmentation       &           \multicolumn{3}{c}{Correlation}           &              \multicolumn{3}{c|}{MAPE}              \\
		&                &                          &      frame      &      video      &     session     &      frame      &      video      &     session     \\
		\hline\hline
		AlexNet         & 224$\times$461 &     None      &     0.6678      &     0.6857      &     0.7371      &     0.1472      &     0.1430      &     0.1368      \\ \hline
		ResNet34        & 224$\times$461 &     None      &     0.6291      &     0.6812      &     0.7398      &     0.1614      &     0.1465      &     0.1339      \\ \hline
		EfficientNet-B0 & 224$\times$461 &     None      &     0.6383      &     0.6739      &     0.7357      &     0.1521      &     0.1399      &     0.1276      \\ \hline
		EfficientNet-B1 & 240$\times$461 &     None      &     0.6735      &     0.7024      &     0.7487           &       0.1428    & 0.1345           &\textbf{ 0.1250} \\ \hline
		EfficientNet-B2 & 260$\times$461 &     None      &     0.6754      &     0.7033      &     0.7630      &     0.1474      &     0.1402      &     0.1298     \\
		\hline\hline
		AlexNet         & 224$\times$461 &     hor. flip      &     0.6587      &     0.6772      &     0.7263      &     0.1474      &     0.1454      &     0.1420      \\ \hline
		ResNet34        & 224$\times$461 &     hor. flip      &     0.6504      &     0.6969      &     0.7556      &     0.1453      &     0.1414      &     0.1355      \\ \hline
		EfficientNet-B0 & 224$\times$461 &     hor. flip      &     0.5857      &     0.6119      &     0.6653      &     0.1653      &     0.1609      &     0.1552      \\ \hline
		EfficientNet-B1 & 240$\times$461 &     hor. flip      &     0.6645      & \textbf{0.6998} &     0.7637      & \textbf{0.1377} & \textbf{0.1338} &     0.1282      \\ \hline
		EfficientNet-B2 & 260$\times$461 &     hor. flip      &     0.6620      &     0.6905      &     0.7576      &     0.1424      &     0.1389      &     0.1317      \\ \hline\hline
		AlexNet         & 224$\times$461 &    hor. flip, $\pm 10^\circ$ rot.     &     0.6644      &     0.6804      &     0.7328      &     0.1506      &     0.1492      &     0.1442      \\ \hline
		ResNet34        & 224$\times$461 &    hor. flip, $\pm 10^\circ$ rot.     &     0.6621      &     0.6963      & \textbf{0.7662} &     0.1428      &     0.1389      &     0.1329      \\ \hline
		EfficientNet-B0 & 224$\times$461 &    hor. flip, $\pm 10^\circ$ rot.     &     0.5994      &     0.6349      &     0.7070      &     0.1669      &     0.1620      &     0.1557      \\ \hline
		EfficientNet-B1 & 240$\times$461 &    hor. flip, $\pm 10^\circ$ rot.     &     0.6642      &     0.6935      &     0.7534      &     0.1448      &     0.1419      &     0.1359      \\ \hline
		EfficientNet-B2 & 260$\times$461 &    hor. flip, $\pm 10^\circ$ rot.     &     0.6661      &     0.6953      &     0.7592      &     0.1441      &     0.1408      &     0.1326      \\ \hline\hline
		AlexNet         & 224$\times$461 & hor. flip, $\pm 10^\circ$ rot., bri./cont. adj & \textbf{0.6695} &     0.6889      &     0.7442      &     0.1498      &     0.1481      &     0.1447      \\ \hline
		ResNet34        & 224$\times$461 & hor. flip, $\pm 10^\circ$ rot., bri./cont. adj &     0.6649      &     0.6959      &     0.7623      &     0.1391      &     0.1349      &     0.1282      \\ \hline
		EfficientNet-B0 & 224$\times$461 & hor. flip, $\pm 10^\circ$ rot., bri./cont. adj &     0.6067      &     0.6471      &     0.7176      &     0.1668      &     0.1619      &     0.1555      \\ \hline
		EfficientNet-B1 & 240$\times$461 & hor. flip, $\pm 10^\circ$ rot., bri./cont. adj &     0.6563      &     0.6847      &     0.7416      &     0.1457      &     0.1429      &     0.1374      \\ \hline
		EfficientNet-B2 & 260$\times$461 & hor. flip, $\pm 10^\circ$ rot., bri./cont. adj &     0.6638      &     0.6916      &     0.7577      &     0.1445      &     0.1415      &     0.1338      \\ \hline
	\end{tabular}
\end{table*}

For what concerns the classification results reported in Tab. \ref{tab:baselines_class},
ResNet outperforms the other CNN architectures on most of the considered metrics.
In particular, EfficientNet-B0 achieves a slightly better accuracy only at frame-level.
As regarding the correlation, ResNet takes full advantage of the strongest augmentation, outperforming the others with a gap of about 2\% with respect to the classification runner-up and 1\% with respect to the best regression result (Tab. \ref{tab:baselines_reg}).

\begin{table*}[!t]
	\renewcommand{\arraystretch}{1.3}
	\caption{Binary classification results of the considered baselines with different augmentations. From top to bottom we report the results without any augmentation and by adding the following operations, consecutively: random horizontal flip, random rotation in the range $\left[-10^\circ, +10^\circ\right]$, random brightness and or contrast variation in the range $\left[-25\%, +25\%\right]$.}
	\label{tab:baselines_class}
	\centering
	\begin{tabular}{|lc|c||ccc|ccc|}
		\hline
		Network         &   Input size   &       Augmentation       &           \multicolumn{3}{c}{Correlation}           &            \multicolumn{3}{c|}{Accuracy}            \\
		&                &                          &      frame      &      video      &     session     &      frame      &      video      &     session     \\ 
		\hline\hline
		AlexNet         & 224$\times$461 &     None      &     0.6586      &     0.6749      &     0.7518      &     0.8121      &     0.8245      &     0.8461      \\ \hline
		ResNet34        & 224$\times$461 &     None      &     0.6507      &     0.6830      &     0.7562      &     0.8129      &     0.8229      &     0.8397      \\ \hline
		EfficientNet-B0 & 224$\times$461 &     None      &     0.6536      &     0.6794      &     0.7463      &     0.8159      &     0.8316      &     0.8461      \\ \hline
		EfficientNet-B1 & 240$\times$461 &     None      &     0.6647      &     0.6928      &     0.7578      &     0.8214      &     0.8332      &     0.8653      \\ \hline
		EfficientNet-B2 & 260$\times$461 &     None      &     0.6471      &     0.6687      &    0.7353      &     0.8166      &     0.8340      &     0.8654      \\ 
		\hline\hline
		
		AlexNet         & 224$\times$461 &     hor. flip      &     0.6593      &     0.6729      &     0.7458      &     0.8054      &     0.8166      &     0.8462      \\ \hline
		ResNet34        & 224$\times$461 &     hor. flip      &     0.6414      &     0.6701      &     0.7526      &     0.8096      &     0.8182      &     0.8397      \\ \hline
		EfficientNet-B0 & 224$\times$461 &     hor. flip      &     0.6425      &     0.6641      &     0.7351      &     0.8141      &     0.8190      &     0.8397      \\ \hline
		EfficientNet-B1 & 240$\times$461 &     hor. flip      &     0.6461      &     0.6752      &     0.7533      &     0.8194      &     0.8356      &     0.8718      \\ \hline
		EfficientNet-B2 & 260$\times$461 &     hor. flip      &     0.6390      &     0.6690      &     0.7431      &     0.8152      &     0.8237      &     0.8397      \\ \hline\hline
		AlexNet         & 224$\times$461 &    hor. flip, $\pm 10^\circ$ rot.     &     0.6485      &     0.6658      &     0.7363      &     0.8124      &     0.8229      &     0.8526      \\ \hline
		ResNet34        & 224$\times$461 &    hor. flip, $\pm 10^\circ$ rot.     &     0.6602      &     0.6869      &     0.7546      &     0.8248      & \textbf{0.8387} & \textbf{0.8782} \\ \hline
		EfficientNet-B0 & 224$\times$461 &    hor. flip, $\pm 10^\circ$ rot.     &     0.6696      &     0.6883      &     0.7543      & \textbf{0.8292} &     0.8364      &     0.8526      \\ \hline
		EfficientNet-B1 & 240$\times$461 &    hor. flip, $\pm 10^\circ$ rot.     &     0.6539      &     0.6794      &     0.7540      &     0.8156      &     0.8356      &     0.8462      \\ \hline
		EfficientNet-B2 & 260$\times$461 &    hor. flip, $\pm 10^\circ$ rot.     &     0.6378      &     0.6634      &     0.7264      &     0.8123      &     0.8245      &     0.8590      \\ \hline\hline
		AlexNet         & 224$\times$461 & hor. flip, $\pm 10^\circ$ rot., bri./cont. adj &     0.6557      &     0.6690      &     0.7310      &     0.8061      &     0.8206      &     0.8590      \\ \hline
		ResNet34        & 224$\times$461 & hor. flip, $\pm 10^\circ$ rot., bri./cont. adj & \textbf{0.6777} & \textbf{0.7030} & \textbf{0.7782} &     0.8208      & \textbf{0.8387} &     0.8526      \\ \hline
		EfficientNet-B0 & 224$\times$461 & hor. flip, $\pm 10^\circ$ rot., bri./cont. adj &     0.6581      &     0.6819      &     0.7586      &     0.8249      &     0.8379      &     0.8590      \\ \hline
		EfficientNet-B1 & 240$\times$461 & hor. flip, $\pm 10^\circ$ rot., bri./cont. adj &     0.6592      &     0.6828      &     0.7564      &     0.8138      &     0.8340      &     0.8590      \\ \hline
		EfficientNet-B2 & 260$\times$461 & hor. flip, $\pm 10^\circ$ rot., bri./cont. adj &     0.6620      &     0.6846      &     0.7501      &     0.8104      &     0.8316      &     0.8718      \\ \hline
	\end{tabular}
\end{table*}

In Tab. \ref{tab:advanced} we report the results obtained by best-performing networks, namely ResNet and EffcientNet-B1, when advanced training strategies are implemented for both regression (top) and classification (bottom) strategies.
Similarly to baseline approaches, the regression results did not improve.
With the considered CNN architectures,
a sort of performance plateau has been reached for this challenging problem,
which would probably benefit from more training samples at the mid-SF range, i.e. between 250 and 350.
On the other hand, the performance increase for classification approaches when preserving the horizontal flip information in the network architecture.
In terms of correlation, the overall ResNet result is further improved reaching the value of 0.7821 at session-level.
As regarding the Curriculum Learning, this positively impacts the accuracy, with EfficientNet-B1 reaching the best result at video-level.

Finally, a comparison with both the visual and automatic approaches pursued in our previous work \cite{raimondi2018visual} has been implemented.
While the textural analysis achieves a Spearman's rank correlation coefficient of 0.6703 at session-level, the visual LUS score assigned by human experts reaches a much higher correlation equals to 0.8259.
A brief comparison with the baseline results confirms the superiority of deep learning approaches over classical handcrafted features, with an improvement in terms of correlation up to 10\%.
At the same time, compared to the visual score of the human experts,
our best result is only 4.4\% worse.

\begin{table*}[!t]
	\renewcommand{\arraystretch}{1.3}
	\caption{Results of the best performing networks with Curriculum Learning and Horizontal position information preserving.}% 1x15 ResNet34, 1x20 EfficientNet-B1.}
	\label{tab:advanced}
	\centering
	\begin{tabular}{|l|l|l||ccc|ccc|}
		\hline
		Train mode & Network            &         Advanced       &      \multicolumn{3}{c}{Correlation}       & \multicolumn{3}{c|}{MAPE or Accuracy} \\
		&                               &         training       & frame  		   &     video      &    session     & frame  		  & video  		   & session\\\hline\hline		 
		%\parbox[t]{2mm}{\multirow{2}{*}{\rotatebox[origin=c]{90}{Reg.}}}
		\multirow{2}{*}{Regression}
		& ResNet34                      &    hor. flip position  & 0.6734 		   &    0.6778      &    0.7440      &0.1681 		  &\textbf{0.1427} &\textbf{0.1344}\\\cline{2-9}
		& EfficientNet-B1               &    hor. flip position  & 0.6361 		   &    0.6754      &    0.7505      &\textbf{0.1511} &0.1475 		   & 0.1395\\\hline\hline
		%\parbox[t]{2mm}{\multirow{4}{*}{\rotatebox[origin=c]{90}{Class.}}}
		\multirow{4}{*}{Classification}
		& ResNet34                      &    hor. flip position  & 0.6766 		   &\textbf{0.7051} &\textbf{0.7821} &0.8179 		  &0.8324 		   &\textbf{0.8718}\\ \cline{2-9}
		& EfficientNet-B1               &    hor. flip position  & 0.6459 		   &    0.6747      &    0.7608      &0.8092 		  &0.8261		   & 0.8333\\\cline{2-9}
		& ResNet34                      &     Curriculum Learning& \textbf{0.6780} &    0.6995      &    0.7604      &\textbf{0.8195} &0.8348 		   & 0.8654\\\cline{2-9}
		& EfficientNet-B1               &     Curriculum Learning& 0.6645 		   &    0.6852      &    0.7600      &0.8161 		  &\textbf{0.8616} & 0.8590\\\hline
	\end{tabular}
\end{table*}

\begin{figure}[!t]
	\centering
	\subfloat[]{\includegraphics[width=2.5in]{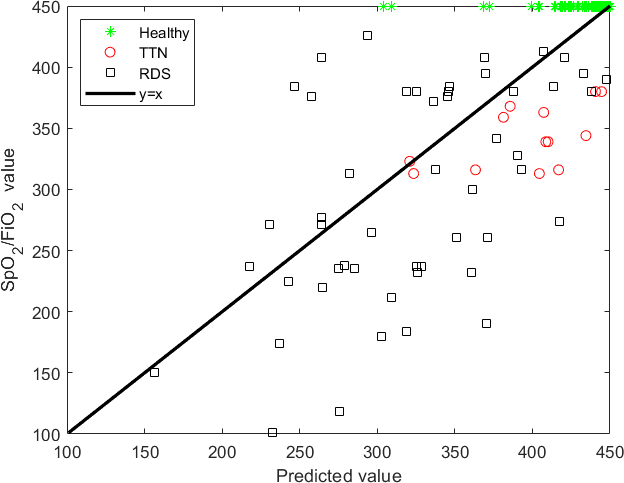}%
		\label{fig:scatterplot_reg_first_case}}
	\hfil
	\subfloat[]{\includegraphics[width=2.5in]{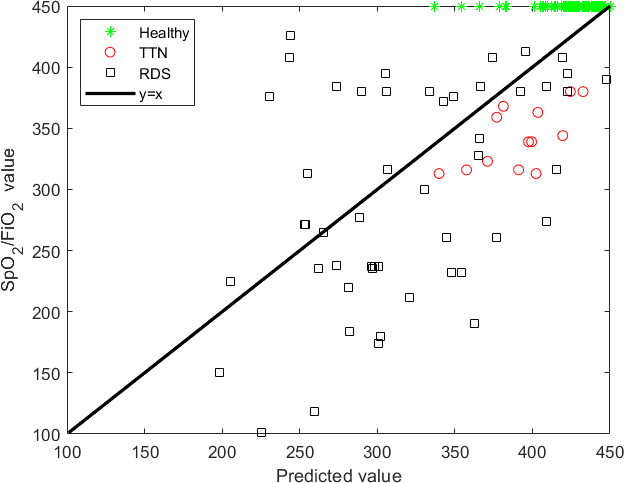}%
		\label{fig:scatterplot_reg_second_case}}
	\caption{Scatter plots of the predicted score with EfficientNet-B1 (top) and ResNet34 (bottom) architectures trained respectively without and with the random rotation.}
	\label{fig:scatterplot_reg}
\end{figure}

\begin{figure}[!t]
	\centering
	\subfloat[]{\includegraphics[width=2.5in]{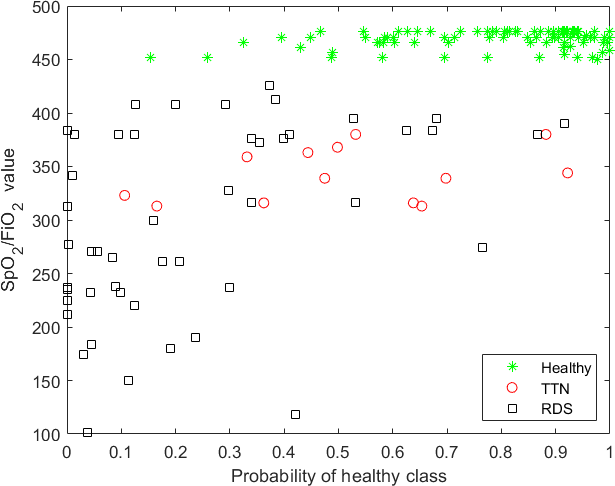}%
		\label{fig:scatterplot_class_first_case}}
	\hfil
	\subfloat[]{\includegraphics[width=2.5in]{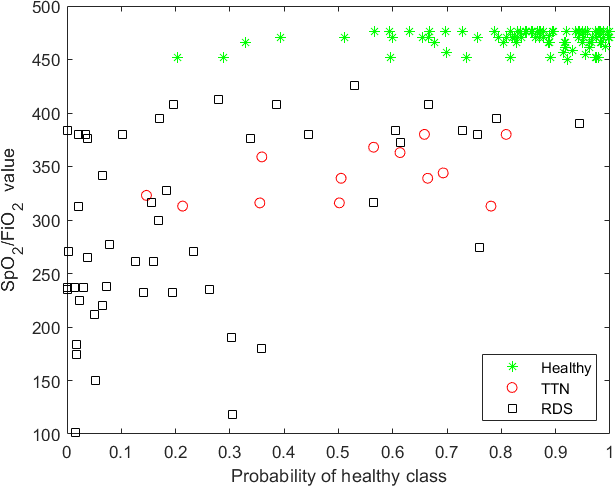}%
		\label{fig:scatterplot_class_second_case}}
	\caption{Scatter plots of the predicted score with ResNet34 architecture trained with classification-based strategy respectively without (top) and with (bottom) brightness/contrast random adjustment.}
	\label{fig:scatterplot_class}
\end{figure}

In order to qualitatively assess the performance, we depict
in Fig. \ref{fig:scatterplot_reg} and \ref{fig:scatterplot_class} the scatter plots of the best performing solution respectively for regression and classification training strategies.
For visualization purposes, all the values in Fig. \ref{fig:scatterplot_reg} are clipped to 450, as it is been done during the training of the networks.
The two figures show the scatter plots for EfficientNet-B1 (a) and ResNet (b).
In both cases, the dispersion around the ideal regression line is higher for mid and lower SF values,
proving the need for more samples in this range.
Often these values, which are responsible for the most correlation performance drop, are associated with RDS patients (black squares).
At the same time, SF value for the TTN patients (red circles) is always overestimated.
This is a very hard task even for the human expert because of the visual similarities between the US of TTN and healthy patients,
which only differs for small details on the pleural line (see Fig. \ref{fig:SchemaImag}).

Moreover, slightly different behavior can be observed in this regard between the two networks.
Indeed, EfficientNet-B1 (a) better predicts a few TTN samples at the cost of errors on the healthy ones,
while ResNet (b) does the exact opposite.
In Fig. \ref{fig:scatterplot_class}, the scatter plots for the best classification results are depicted, namely ResNet either without (a) or with (b) the brightness/contrast augmentation.
Obviously, in this case, we do not expect a linear correlation with the target marker.
With respect to the regression results, the RDS samples are more concentrated on the low part of the ranges,
with some outliers mostly concentrated in the SF range between 300 and 400.
When a stronger augmentation is performed (b), the healthy and TTN patients seem better concentrated respectively at the mid and high predicted range.

\section{Conclusion}
\label{sec:conclusions}
In this work, we presented a non-invasive experimental analysis about the use of deep learning techniques to predict the lung respiratory status starting from its ultrasound image.
To this aim, a large dataset of 87 patients and about 1200 US videos has been collected from 3 different centers.
Experimental results show that ResNet34 trained for binary classification achieves the best performance in terms of correlation with the selected reference marker.
Moreover, by modifying the CNN architecture in order to take into account the horizontal position of the extracted convolutional networks, the correlation further improves.
It is worth observing that the proposed approach performs comparably with the human operator.

Future research will be devoted to enlarge the dataset including data from other medical centers and improve the training strategy by exploiting the temporal information of the LUS videos. 

\bibliographystyle{IEEEtran}
\bibliography{IEEEabrv,bibliography}
\end{document}